# Local field correction effects on quasi-particle inelastic scattering rate in a coupled-quantum-layers system at finite temperature

Vahdat Rafee[*], Alireza Razeghizadeh, Abdolrasoul Gharaati

*Department of Physics, Payamenoor University, Tehran P.O. BOX19395-3697, IRAN*

[*]Corresponding author email: V.Rafee@gmail.com

**Abstract**

Local field correction effects on intra-layer inelastic scattering rate of interacting electrons are theoretically investigated in a coupled-quantum-wells structure, at finite temperature. At first, temperature dependent dynamic dielectric function is calculated using random phase approximation (*RPA*). Then, local field correction effects are considered in calculations by employing short-range effects of exchange-correlation holes around electrons. We employ Hubbard, finite-temperature Hubbard and *STLS* approximations. Finally, quasi-particle inelastic scattering rate is calculated using the imaginary part of the electron self-energy within *GW* method. The results show that quasi-particle inelastic scattering rate is reduced, when a local field correction is employed, at any temperature, wave vector and electron density.

## 1. Introduction

In recent decades, many scientists have focused on nanoscale systems. The coupled-quantum-layers is one of the most important systems and a lot of theoretical and experimental studies have been done on this interesting structure [1-22].

In fact, a coupled-quantum-layers structure is composed of two parallel semiconductor nano-layers which are separated from each other by few nanometers distance. The potential between the two layers is high enough to prevent electron tunneling.

These bilayer structures are usually made of different nanolayers materials such as *GaAs*, *GaAlAs*, graphene, phosphorene, Silicene, germanene. In recent years, many interesting works are done on the different bilayers such as graphene [12-14, 23-27], phosphorene [18], Silicene [27-29], and germanene [29].

These systems are studied in recent works because these systems have many interesting physical properties such as the inelastic lifetimes [24, 30-34], Coulomb drag [6, 13, 14, 21], electron mobility [8, 16, 20, 24], the energy transfer rate [15, 17, 19], the inelastic Coulomb scattering rate [7, 22, 24, 30, 33, 35-37], etc.

One of the considered properties in recent researches is the quasi-particle inelastic scattering rate and the inelastic lifetime of interacting electrons [4, 7, 22, 24, 30-34, 36-39]. The quasi-particles inelastic scattering rate gives a measure on the average time interval between two consecutive scatterings in an interacting electron system due to Coulomb interaction. The physical properties such as transport and the rate of tunneling, localization are studied easily using calculation the quasi-particles inelastic scattering rate.

In the last two decades, many theoretically and experimentally works are done on the quasi-particles inelastic scattering rate of *2DEG* such as geometry effect [31, 35], local field correction effect in finite temperature [33], Hubbard correction infinite temperature [36, 40] and, Coulomb scattering lifetime [30]. Also, interesting researches are done on this parameter of *DQW* such as electron-electron scattering [7], Coulomb scattering lifetime [24], Coulomb drag[6, 13, 14, 21] and geometry effect in finite temperature [17, 22, 41].

The dynamical dielectric function is the most important parameter for calculation of quasi-particles inelastic scattering rate. This physical quantity of many-body particle systems such as *2DEG* is calculated using different approximations such as Tomas-Fermi, *RPA*, Hubbard, temperature-dependent Hubbard and *STLS* [1, 3, 40].

One of the most important approximations for calculation dynamical dielectric function of a many-body system is the random phase approximation (*RPA*). This approximation ignores exchange-correlation holes around electrons. This approximation is reliable for high electron density systems in which short-range effects do not matter. The short-range interaction is really important when the electron density is low. Therefore, the local field correction is used in the calculation of dynamical dielectric function using Hubbard, temperature dependence Hubbard [33, 35, 36, 42] and Singwi, Tosi, Land, and Sjölander (*STLS*) approximation [5, 33]. These approximations are considered exchange-correlation effects in the calculations.

In the present paper, we study the quasi-particles inelastic scattering rate due to Coulomb electron-electron interaction for a *DQW*. At first, we calculate a dynamic dielectric function of the system by making use of the *RPA* approximation. Then, we employ local field corrections implemented in Hubbard, temperature-dependence Hubbard and *STLS* approximations. The quasi-particle inelastic scattering rate is calculated using *GW* approximation. The *GW* approximation has been presented by Zhang and Das Sarma [30].

The rest of the paper is structured as follows: in section 2, we present a theoretical formalism for studying the quasi-particles inelastic scattering rate and *GW* approximation. In section 3 we describe and discuss our numerical results. Finally, in section 4, we summarized highlights of this work.

## 2- Theoretical formalism

We consider two coupled parallel *n*-doped *GaAs* semiconductor nanolayers with *L* thickness which are separated from each other by *d*. The temperature of the system is considered to be low enough so that electron- phonon interaction can be neglected. We consider the electron density of the two layers to be the same in the calculations, therefore the scattering rate does not change with changing the sub-band index of layers.

We use atomic units in the calculations ($\hbar = e^2/2m^* = 1$) [1]. The fermi energy, wave vector, and temperature are defined as $T_f = E_f/k_B, E_f = \hbar^2 K_f^2/2m^*$, and $K_f = \sqrt{2}/a_B^* r_s$ for 2DEG, respectively[1]. Where $k_B$ is Boltzmann constant and $a_B^* = (\varepsilon_s/m^* e^2)$ is the effective Bohre reduce ($m^*$ is the effective mass and $\varepsilon_s$ is the background dielectric constant) and $r_s = 1/(a_B^* \sqrt{\pi n})$ is the dimensionless density parameter which determines the average distance between the interacting electron in the system ($n = K_f^2/2\pi$ is the electron density of each layer).

At first, the dynamic dielectric function is calculated within *RPA* approximation which is reliable at high electron density or equivalently low dimensionless density parameter $r_s \leq 1$, long-range interactions are considered and short-range interactions have neglected. The results of *RPA* approximation are known to be not accurate enough when the electron density decreases, $r_s > 1$. In the mentioned condition, the short-range interactions due to exchange-correlation holes must be taken into account through local field corrections such as Hubbard, temperature-dependent Hubbard and *STLS* approximation.

**2-1-RPA approximation**

The dynamic dielectric function of *2DEG* in *RPA* approximation is written as [1]:

$$\varepsilon(\mathbf{q},\omega) = 1 - \frac{V(\mathbf{q})\chi^0(\mathbf{q},\omega)}{1 + V(\mathbf{q})\chi^0(\mathbf{q},\omega)G_H(\mathbf{q})} \tag{1}$$

Here, $V(\mathbf{q})$ is the *2D* bare Coulomb potential and $\chi^0(\mathbf{q},\omega)$ is the non-interacting electronic density-density response function and $G_H(\mathbf{q})$ is the local field correction function.

In *RPA* approximation $G_H(\mathbf{q})$ is replaced by zero, therefore, the non-interacting density-density response of *2DEG* is given as [1]:

$$\chi_{2D}(\mathbf{q},\omega) = -\frac{1}{\pi\hbar}\int d^2\mathbf{K}\, n_k(\varepsilon)\left[\frac{1}{\omega - \left(\mathbf{q}\cdot\mathbf{K} + \frac{q^2}{2}\right) + i\eta} - \frac{1}{\omega + \left(\mathbf{q}\cdot\mathbf{K} + \frac{q^2}{2}\right) + i\eta}\right] \tag{2}$$

where $n_k = 1/(\exp[(\varepsilon - \mu)/k_B T] + 1)$ is the Fermi-Dirac distribution function at non-zero temperature equilibrium system.

In order to simplify the calculation, we define the dimensionless quantities as follows [6]:

$$t \equiv \frac{K_B T}{E_F},\, K = \frac{k}{k_F},\, \tilde{\mu} \equiv \frac{\mu}{E_F},\, E_F \equiv \mu(T=0),\, Q \equiv \frac{q}{k_F},\, \Omega \equiv \frac{\hbar\omega}{E_F},\, \tilde{\chi} \equiv \frac{\chi^{(0)}}{(m/\pi\hbar^2)} \tag{3}$$

Therefore, the non-interacting density-density response function of 2DEG is rewritten as [6]:

$$\tilde{\chi}_{2D}(Q,\Omega) = \frac{1}{(2\pi)}\int dk_x \int dk_y\, n(k_x,k_y)\left[\frac{1}{\frac{\Omega}{2} - (Qk_x + \frac{Q^2}{2}) + i\eta} - \frac{1}{\frac{\Omega}{2} + (Qk_x + \frac{Q^2}{2}) + i\eta}\right] \tag{4}$$

At finite temperature, the imaginary part of the non-interacting density-density response of *2DEG* is given as [6]:

$$\text{Im}[\chi_{2D}(\mathbf{q},\omega)] = N_0 \frac{\sqrt{t}\pi}{2Q}\left[F_{-\frac{1}{2}}\left(\frac{A^+}{t}\right) - F_{-\frac{1}{2}}\left(\frac{A^-}{t}\right)\right] \tag{5}$$

where $A^\pm = \tilde{\mu} - (\Omega/2Q \pm Q/2)^2$, $\tilde{\mu} = t[\ln(1/t) - 1]/k_B T$, $N_0 = m^*/\pi\hbar$ and $F_{-1/2}$ is the Fermi function of order $-1/2$ is given as [6]:

$$F_{-\frac{1}{2}}(x) = \frac{1}{\sqrt{\pi}}\int_0^\infty dy\, \frac{y^{-\frac{1}{2}}}{e^{(y-x)} + 1} \tag{6}$$

Similarly, the real part of the non-interacting density-density response of *2DEG* is given as [6]:

$$\operatorname{Re}[\chi_{2D}(\mathbf{q},\omega)] = -\frac{N_0}{\exp(-\tilde{\mu}/t)+1} + \frac{\operatorname{sgn}(a_+)}{Q} M_t(a_+^2) - \frac{\operatorname{sgn}(a_-)}{Q} M_t(a_-^2) \qquad (7)$$

where $M_t(X) = \int_0^\infty [(X-\mu')^{1/2}/4t \cosh^2(\tilde{\mu}'-\tilde{\mu}/2t)] d\tilde{\mu}'$ and **sgn** is the Sign function,

$a_\pm = (\Omega/Q \pm Q)/2$ [6].

The dynamical dielectric function of *DQW* in *RPA* approximation is written as [5]:

$$\varepsilon(\mathbf{q},\omega,T) = \begin{pmatrix} 1-V_{11}(\mathbf{q})\chi_1(\mathbf{q},\omega,T) & V_{12}(\mathbf{q})\chi_1(\mathbf{q},\omega,T) \\ V_{21}(\mathbf{q})\chi_2(\mathbf{q},\omega,T) & 1-V_{22}(\mathbf{q})\chi_2(\mathbf{q},\omega,T) \end{pmatrix} \qquad (8)$$

where $V_{ii}$ is the unscreened electron-electron inter-layer Coulomb potential and $V_{ij}$ is the unscreened electron-electron Coulomb interaction matrix elements in the atomic units is written as [6, 22]:

$$V_{ij}(q) = \frac{2\pi e^2}{q\varepsilon_s} F_{ij}(q) \exp(-qd(1-\delta_{ij})) \qquad (9)$$

In the above equation $F_{ij}(\vec{q})$ is the form factor which for an infinite rectangular quantum well is given as[6, 22]:

$$F_{ii}(X) = \frac{3X + \frac{8\pi^2}{X}}{X^2 + 4\pi^2} - \frac{32\pi^4[1-\exp(-X)]}{X^2(X^2+4\pi^2)^2} \qquad (10)$$

$$F_{ij}(X) = \frac{64\pi^4 \sinh^2\left(\frac{X}{2}\right)}{X^2(X^2+4\pi^2)^2} \exp(-qd) \qquad (11)$$

Here, $d$ is the distance between the two layers and $L$ is the thickness of the layers and $X = qL$.

**2-2- Temperature-dependent and temperature-independent Hubbard approximations:**
The exchange hole and short-range effect without considered temperature effect in the calculation of dynamic dielectric function of *2DEG* is studied using zero temperate Hubbard approximation [33, 35, 36, 42].
In the calculation of dynamical dielectric function, we define $G_H(\mathbf{q})$ the *2D* local field correction in Hubbard approximation as fallows[1]:

$$G_H(\mathbf{q}) = \frac{1}{2} \frac{q}{\sqrt{q^2+k_F^2}} \qquad (12)$$

In order to consider temperature effects in short-range interactions, in the calculations of the dynamic dielectric function of *2DEG,* we use temperate-dependent Hubbard approximation [15, 17, 21, 36].
Also, in temperature-dependent Hubbard approximation $G_H(\mathbf{q})$ is defined using $G_H(\mathbf{q},T)$, in which the Fermi energy is replaced by temperature-dependent chemical potential [1].

$$G_H(\mathbf{q},T) = \frac{1}{2}\frac{q}{\sqrt{q^2 + 2m^*\mu(T)/\hbar^2}} \qquad (13)$$

**2-3- STLS approximation:**
In order to consider exchange-correlation hole and also short-range effect, in the calculation of dynamic dielectric function of *2DEG*, we use *STLS* approximation [33, 43].
In this approximation $G_H(\mathbf{q})$ is replaced by $G_{STLS}(\mathbf{q})$ which the *2D* local field correction in *STLS* approximation is given as [33]:

$$G_{STLS}(\mathbf{q}) = -\frac{1}{n}\int \frac{d\mathbf{q}'}{(2\pi)^2}\frac{\mathbf{q}.\mathbf{q}'}{q^2}\frac{v(\mathbf{q}')}{v(\mathbf{qq})}[S(|\mathbf{q}-\mathbf{q}'|)-1] \qquad (14)$$

$$S(\mathbf{q}) = \frac{-\hbar}{n\pi v(\mathbf{q})}\int_0^\infty \mathrm{Im}[\varepsilon^{-1}(\mathbf{q},\omega)] \qquad (15)$$

where $v(\mathbf{q})$ is the Fourier transform of Coulomb interaction for electrons in the lowest subband. Gold and Calmes have been calculated G for *2DEG* as [44]:

$$G_{STLS}(y) = r_s^{2/3}\frac{1.402 y}{[2.644 C_{12}^2 + y^2 C_{22}^2]^{1/2}} \qquad (16)$$

Here, $y = q/q_0$, $q_0 = 2/(r_s^{2/3} a_B^*)$, $C_{12} = 1.914 r_s^{0.119}$, $C_{22} = 1.640 r_s^{0.530}$.

In order to consider short-range interaction and local field correction, in the calculation of dynamic dielectric function of *DQW* $\varepsilon(\mathbf{q},\omega,T)$ in Hubbard, temperature-dependent Hubbard, and *STLS* approximations, we neglect the short-range interaction on inter-layer interactions and replace the intra-layer potential $V_{ii}$ with $V_{ii}(1-G(\mathbf{q}))$ [19].

**2-4- The quasi-particle inelastic scattering rate**
The quasi-particle inelastic scattering rate of *DQW*, $\Gamma$, or equivalently reverse of the inelastic scattering lifetime, $\tau^{-1}$ is calculated using the imaginary part of the electron self-energy, $\Sigma$, as follows [1]:

$$\Gamma_{ii}(\mathbf{k},\zeta_k,T) = \tau_{ii}^{-1}(\mathbf{k},\zeta_k,T) = -2\,\mathrm{Im}\,\Sigma_{ii}(\mathbf{k},\zeta_k,T) \qquad (17)$$

where **k** is momentum and $\zeta_k = k^2/2m^* - E_f$ is unscreened energy.
The electron self-energy in *GW* method is calculated using the screened interaction, *W*, and the dressed Green function. The *GW* calculations are a difficult self-consistent problem to solve. These calculations are much simpler when non-interaction Green function replaces *G* in theory.
In the Matsubara formalism the electron self-energy is given as follows [1]:

$$\Sigma_{ii}(\mathbf{k},\omega_n) = -\frac{1}{\beta V}\sum_{\mathbf{q}}\sum_{\nu_n} W_{ii}(\mathbf{q},\nu_n) G_{ii}^0(\mathbf{k}+\mathbf{q},\omega_n+\nu_n) \qquad (18)$$

where $\nu_n = 2n\pi/\beta\hbar$, $\omega_n = (2n+1)\pi/\beta\hbar$ and *V* is the volume of the electron gas, $\beta = (k_B T)^{-1}$, *i* is index layer and *n* is an integer number.
The non-interacting Green function is written as [1]:

$$G^0(\mathbf{k},\omega_n) = 1/[i\omega_n - \hbar^{-1}(E_\mathbf{k} - \mu)] \tag{19}$$

The dynamically screened intra-layer electron-electron interaction can be obtained as[5]:

$$W_{ii}(q,\omega,T) = \frac{\left(1 - v_{jj}(q)\chi_{jj}^0(q,\omega,T)\right)v_{jj}(q) + v_{ji}^2(q)\chi_{ii}^0(q,\omega,T)}{\det[\varepsilon(q,\omega,T)]} \tag{20}$$

Finally, the imaginary part of electron self-energy, $\Sigma$, is given as [30]:

$$\operatorname{Im}\Sigma_{ii}(\mathbf{k},\xi_k,T) = \frac{1}{V}\sum_{\mathbf{q}}\operatorname{Im}\left(W_{ii}(q,\zeta_{\mathbf{q}+\mathbf{k}} - \zeta_k,T)\right) \times \left(f_B(\zeta_{\mathbf{k}+\mathbf{q}} - \zeta_k,T) + f_F(\zeta_{\mathbf{k}+\mathbf{q}},T)\right) \tag{21}$$

Where $f_F$ and $f_B$ are Fermi-Dirac and Bose-Einstein distribution functions, respectively.

**3- Result and discussion**
In this section, numerical results of the local field corrections on the quasi-particle inelastic scattering rate of *DQW* are presented. Quasi-particles, acoustic plasmons, and optical plasmons have all contributions in the inelastic Coulomb scattering rate for an interacting electron system [22].
in recent researches, the calculation of the inelastic Coulomb scattering rate of *DQW* is showed that the quasi-particles, acoustic plasmons, and optical plasmons are played role in the inelastic Coulomb scattering rate, respectively with increasing the value of k dimensionless wave vectors [22].
The system studied here is an *n*-doped *GaAs*-based *DQW* which is modeled as an infinite double rectangular quantum wells structure with 50nm wells separation (*d=50nm*) and well thickness of *20nm* (*L=20nm*).

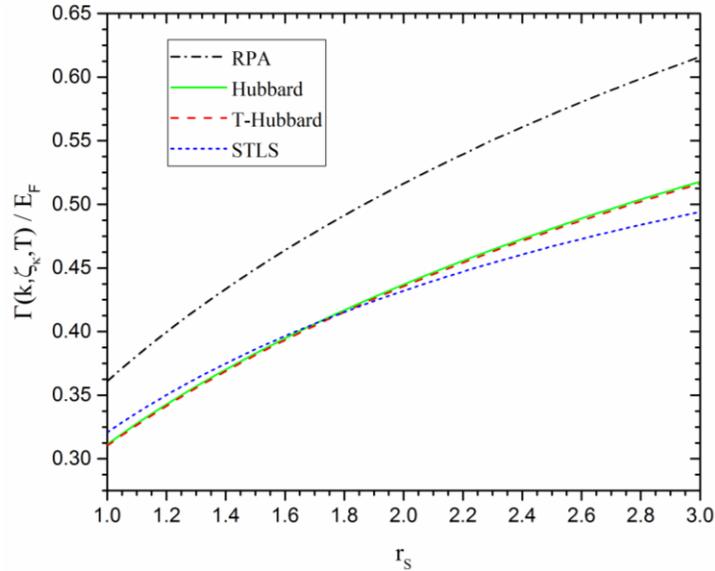

**Fig.1.** The quasi-particle inelastic scattering rate of DQW structure in RPA, Hubbard, temperature-dependent Hubbard and STLS approximation (in Fermi energy unit) as function of density parameter
$(T = 0.5T_f, L = 20nm, d = 50nm, k = 1.4k_f)$

Fig.1 shows the quasi-particle inelastic scattering rate versus the dimensionless electron density parameter within *RPA*, Hubbard, temperate-dependent Hubbard and *STLS* approximations, where the used parameters in the numerical calculation for wave vector, thickness, separation, and temperature are $k=1.4k_f$, $L=20nm$, $d=50nm$, $T=0.5T_f$, respectively.

In addition, Fig.1 shows that the quasi-particle inelastic scattering rate increases with increasing dimensionless electron density parameter. Also, the behavior of the quasi-particle inelastic scattering rate in *STLS* approximation is changed at $r_s=1.8$, so that quasi-particles inelastic scattering rate has higher values within STLS approximation compared to Hubbard and temperature-dependent Hubbard for $r_s \leq 1.8$, and lower values for $r_s > 1.8$.

Another important point in Fig. 1 is that the results of the quasi-particle inelastic scattering rate in the local field correction are far away from obtained results within *RPA*. This shows that exchange-correlation effect gets more and more important with increasing dimensionless electron density parameter.

Moreover, Fig.1 shows that the quasi-particle inelastic scattering rate within temperature-dependent Hubbard has lower values compared with Hubbard approximation at all of the dimensionless electron density parameters studied herein.

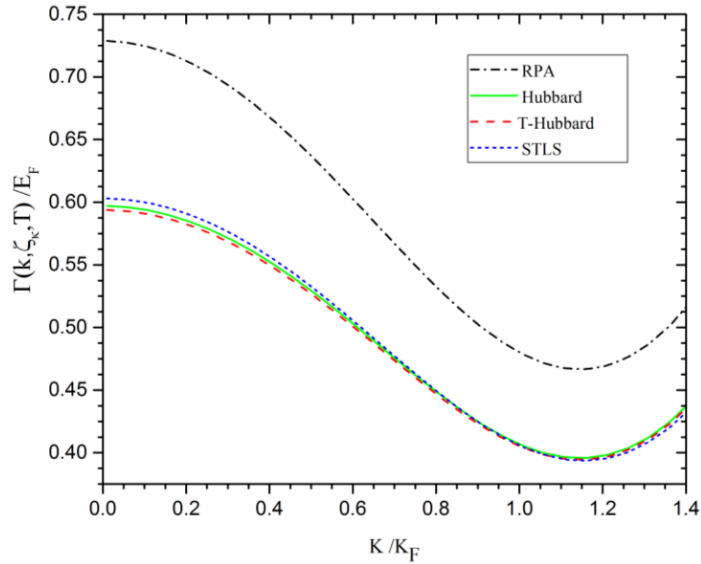

**Fig.2.** The quasi-particle inelastic scattering rate of DQW structure in RPA, Hubbard, temperature-dependent Hubbard and STLS approximation (in Fermi energy unit) as function of dimensionless wave vector
$(T=0.5T_f, L=20nm, d=50nm, r_s=2)$

Fig.2 depicts the quasi-particle inelastic scattering rate versus dimensionless wave vector calculated within *RPA*, Hubbard, temperate-dependent Hubbard and *STLS* approximations, where the used parameters in the numerical calculations are $r_s=2$, $L=20nm$, $d=50nm$, $T=0.5T_f$.

According to Fig.2, we learn that like what we observed in Fig.1, the behavior of the quasi-particle inelastic scattering rate in *STLS* approximation is changed at $k/k_f=0.85$. Also, the quasi-particle inelastic scattering rate in temperature-dependent Hubbard has lower values versus Hubbard approximation at all of the studied dimensionless wave vectors. Furthermore, similar to the previous figure, the results of the quasi-particle inelastic scattering rate in the local field correction far away from obtained results of the *RPA* approximation.

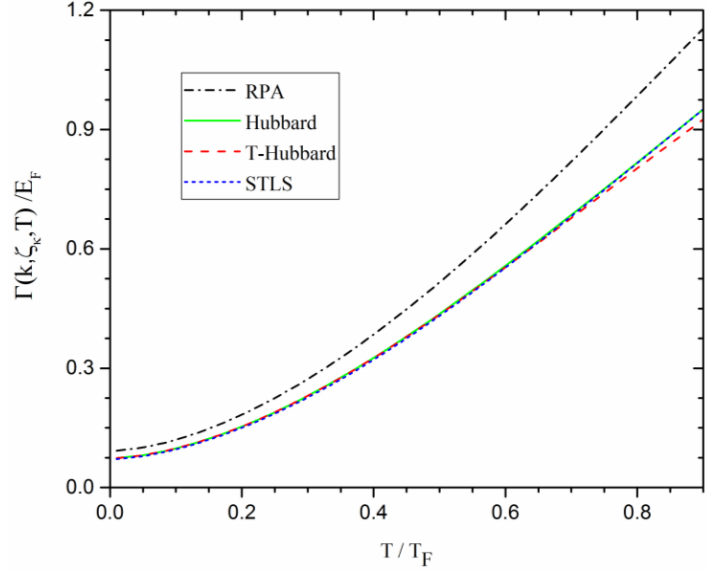

**Fig.3.** The quasi-particle inelastic scattering rate of DQW structure in RPA, Hubbard, temperature-dependent Hubbard and STLS approximation (in Fermi energy unit) as function of the dimensionless temperature
( $k = 1.4k_f, L = 20nm, d = 50nm, r_s = 2$ )

Fig.3 shows the calculated quasi-particle inelastic scattering rate versus dimensionless temperature employing *RPA*, Hubbard, temperate-dependent Hubbard and *STLS* approximations, where the used parameters in the numerical calculations for the dimensionless electron density parameter, thickness, separation, and wave vector are $r_s = 2, L = 20nm, d = 50nm, k = 1.4T_f$ , respectively.

The results of Fig.3 show that the quasi-particle inelastic scattering rate increases with increasing dimensionless temperature. Also, results of local field correction have lower values at all the dimensionless temperatures. In addition, the calculation of quasi-particle inelastic scattering rate within Hubbard, temperature-dependent Hubbard and *STLS* result to similar values at low temperature but with increased temperature, the quasi-particle inelastic scattering rate of temperature-dependent Hubbard have a lower value versus other local field approximation.

### 4- Conclusion

In the paper, we investigated the effect of local field correction on the quasi-particles inelastic scattering rate within an *n*-type Coulomb-coupled double-layer *GaAs*-based structure at different temperatures, wave vectors, and electron density parameters.

The quasi-particle inelastic scattering rate is studied using the calculation of imaginary part of electron self-energy in $G^0W$ approximation and employing *STLS*, Hubbard, temperature-dependent Hubbard, and *RPA* approximations. Comparing the results shows that the calculation of quasi-particle inelastic scattering rate within *RPA*, Hubbard, temperature-dependent Hubbard, and *STLS* result to similar values, at low temperate and low electron density parameter. Also, the quasi-particle inelastic scattering rate within temperature-dependent Hubbard usually have less value versus Hubbard approximation.

### 5- Reference